\documentclass[aps,preprint,groupedaddress]{revtex4}
\usepackage[dvips]{graphicx}
\usepackage{amsmath}
\usepackage{graphicx}
\usepackage{amsfonts}
\usepackage{amssymb}
\usepackage{epsfig}
\usepackage{subfigure}
\usepackage[usenames]{color}

\newcommand{\be}{\begin{equation}}
\newcommand{\ee}{\end{equation}}
\newcommand{\bea}{\begin{eqnarray}}
\newcommand{\eea}{\end{eqnarray}}
\newcommand{\ba}{\begin{array}}
\newcommand{\ea}{\end{array}}
\newcommand{\beas}{\begin{eqnarray*}}
\newcommand{\eeas}{\end{eqnarray*}}
\newcommand{\bes}{\begin{equation*}}
\newcommand{\ees}{\end{equation*}}

\def\i2           {\mbox{$\frac{i}{2}$}}

\begin{document}
\bibliographystyle{apsrev}

\preprint{CERN-PH-TH/2011-113}

\title{\bf Magnetized Domain Walls in the Deconfined Sakai-Sugimoto Model at Finite Baryon Density}

\author{Piyabut Burikham$^{1,2,3}$}\email{Email:piyabut@gmail.com, piyabut.b@chula.ac.th}
\affiliation{$^1$Theoretical High-Energy Physics and Cosmology
Group,
Department of Physics, \\
Faculty of Science, Chulalongkorn University, Bangkok 10330,
Thailand.}
\affiliation{$^2$Thailand Center of Excellence in
Physics, CHE, Ministry of Education, Bangkok 10400, Thailand.}
\affiliation{$^3$Department of Physics, CERN Theory Division CH-1211
Geneva 23, Switzerland.}
\author{Tossaporn Chullaphan$^{1}$}\email{Email:chullaphan.t@gmail.com}
\affiliation{$^1$Theoretical High-Energy Physics and Cosmology
Group,
Department of Physics,\\
Faculty of Science, Chulalongkorn University, Bangkok 10330,
Thailand.}

\begin{abstract}
\noindent  \\

The magnetized pure pion gradient~($\mathcal{5}\varphi$) phase in
the deconfined Sakai-Sugimoto model is explored at zero and finite
temperature.  We found that the temperature has very small effects
on the phase.  The thermodynamical properties of the phase shows
that the excitations behave like a scalar solitonic free particles.
By comparing the free energy of the pion gradient phase to the
competing multiquark-pion gradient~(MQ-$\mathcal{5}\varphi$) phase,
it becomes apparent that the pure pion gradient is less
thermodynamically preferred than the MQ-$\mathcal{5}\varphi$ phase.
However, in the parameter space where the baryonic chemical
potential is smaller than the onset value of the multiquark, the
dominating magnetized nuclear matter is the pion gradient phase.

\end{abstract}

\maketitle

\newpage
\section{Introduction}

Physics of dense nuclear matter is one of the most challenging area
due to the lack of appropriate theoretical modeling.  On one hand,
the entities in the nuclear matter strongly couple to one another
and therefore the perturbative treatment cannot be applied in a
straightforward manner.  On the other hand, the lattice approach to
the Quantum Chromodynamics~(QCD) can be applied to the situations of
hot nuclear matter.  The lattice results predict the deconfinement
phase transition at temperature around 175 MeV for the dilute
nuclear matter.  However, this approach also faces difficulty in
describing the nuclear matter with finite density due to the fermion
sign problem.

An alternative and complementary approach is the application of the
holographic principle or the variations of the AdS/CFT
correspondence~\cite{maldacena,wit0,agmoo} in string theory to study
the behaviour of nuclear matter.  The Sakai-Sugimoto~(SS)
model~\cite{ss lowE, ss more} is a holographic model which mimics
the QCD at low energy most accurately.  Starting with a type IIA
string background with D4-branes as the source.  Take the
near-horizon limit and add the black hole horizon to generate
Hawking-Page temperature to be identified with the temperature of
the dual gauge matter.  Since we need an approximately 4 dimensional
QCD, one of the 5 dimensional subspace is compactified into a circle
whose radius is chosen so small that the Kaluza-Klein states are
much heavier than the relevant energy scales and temperatures.

The quarks and antiquarks are introduced as open-string excitations
on the stack of $U(N_{f})$ flavour D8 and $\overline{\text{D8}}$
branes located at fixed separation distance in the compactified
coordinate. The boundary conditions of the sparticles in the circle
are chosen to be antisymmetric at the location of the flavour branes
and the zeroth modes are thus eliminated. Consequently, the gauge
theory at the flavour branes is a SUSY-broken 5 dimensional
Yang-Mills theory with quarks and antiquarks in the fundamental
representation.  The effective theory has the same particle content
as the QCD.  Using the AdS/CFT correspondence, the bulk theory of
this brane configuration is conjectured to be dual to the QCD-like
gauge theory at the boundary. The striking feature of the SS model
is that it provides a natural geometric realization of the chiral
symmetry breaking.  When the D8 and $\overline{\text{D8}}$ merge at
certain location in the radial coordinate, the quarks and antiquarks
do not transform independently under the chiral transformation and
therefore the chiral symmetry is broken in the connected brane
configuration.  A chiral symmetric configuration occurs when the two
flavour branes are parallel and the dual gauge matter will be in the
chiral symmetric phase.

Subsequent investigation reveals that the SS model accommodates the
exotic possibility that the chiral symmetry restoration and the
deconfinement can occur separately~\cite{asy} when the distance
between the D8 and $\overline{\text{D8}}$ branes in the compactified
dimension is not too large. The deconfinement could occur at
relatively low temperature while the chiral symmetry would be
restored at larger temperature. Even though both the chiral symmetry
breaking and the confinement are results of the strong coupling of
the gauge theory, they are independent of one another as far as we
know.  It is thus possible that the real QCD also has distinctive
chiral symmetry restoration and deconfinement.

Chiral condensate of the QCD-like dense matter is explored in
Ref.~\cite{sst} using the Wess-Zumino-Witten induced anomalous term
in the chiral perturbation theory and in Ref.~\cite{st} using the
bottom-up AdS/QCD based on the confined SS model. When the magnetic
field is applied, the condensate will respond by developing a
gradient in the direction of the applied field. This gradient also
carries the baryonic charge density proportional to the applied
field and the gradient of the condensate. Holographic studies of the
chiral condensate response to the magnetic field is investigated in
Ref.~\cite{BLLm} for the confined SS model.  In Ref.~\cite{prs}, the
pure pion gradient phase is explored and compared with the chiral
symmetric quark-gluon plasma phase in the zero temperature
approximation of the deconfined SS model.  In Ref.~\cite{pbm}, it is
roughly compared with the mixed phase of the multiquark-pion
gradient~(MQ-$\mathcal{5}\varphi$) using a zero-instanton limit of
the multiquark configuration. The preliminary results suggest that
the pure pion gradient phase might be thermodynamically less
preferred than the MQ-$\mathcal{5}\varphi$ phase.  In this article,
we perform a thorough investigation into the pure pion gradient
phase at finite temperature as well as its thermodynamical
comparison to the MQ-$\mathcal{5}\varphi$ in order to obtain a more
definitive quantitative result.  It is found that the pure pion
gradient phase is insensitive to the change of temperature in the
range $T=0-0.16$. It is also shown that the pure pion gradient phase
is generically less preferred than the MQ-$\mathcal{5}\varphi$ phase
except when the baryon chemical potential is smaller than the onset
value of the multiquarks.  In that region of the phase diagram, the
dominating phase is the pure pion gradient.

The article is organized as the following.  In Section II, we setup
the holographic model of the magnetized chirally broken nuclear
phase without an instanton.  A zero temperature solution is obtained
and relevant dual physical quantities as well as their relationships
are discussed. Thermodynamical properties of the pure pion gradient
phase at finite temperature and the comparison with the multiquark
phase are discussed in Section III. Section IV concludes the
article.

\section{Holographic setup of the magnetized chirally broken phase }

In the non-antipodal SS model, a stack of $N_{c}$ D4-branes
generates a curved 10 dimensional spacetime in type IIA string
theory.  The near-horizon limit of this background is then taken and
the black hole horizon is added by introducing the factor
$f(u)$~\cite{wit,agmoo} into the background.  The $x^{4}$ direction
is compactified with certain radius to obtain an effective $(1+3)$
dimensional subspace in the low energy limit.  The resulting
spacetime of the Sakai-Sugimoto model is in the form
\begin{equation}
ds^2=\left( \frac{u}{R_{D4}}\right)^{3/2}\left( f(u) dt^2 +
\delta_{ij} dx^{i}
dx^{j}+{dx_4}^2\right)+\left(\frac{R_{D4}}{u}\right)^{3/2}\left(u^2
d\Omega_4^2 + \frac{du^2}{f(u)}\right)\\ \nonumber
\end{equation}

\begin{equation}
e^{\phi}=g_s \left( \frac{u}{R_{D4}}\right)^{3/4} ,\quad\quad
R_{D4}^3\equiv \pi g_s N_{c} l_{s}^3,\nonumber
\end{equation}

\noindent where $f(u)\equiv 1-u_{T}^{3}/u^3$, $u_T=16{\pi}^2
R_{\text{D4}}^3 {T^2} /9$.  $T$ is the Hawking-Page temperature of
the black hole which is identified with the temperature of the dual
gauge matter at the boundary.  $R_{D4}$ is the curvature of the
background which is generically different from the compactified
radius $R$ of the $x^{4}$ coordinate.  $\phi$ is the dilaton field,
a function of $u$ in this background.

We then introduce stacks of $N_{f}$ D8 and $\overline{\text{D8}}$
flavour branes with separation $L$ on the circle of compactified
$x^{4}$ at the boundary $u\to \infty$.  Open string excitations with
one end on these branes behave like chiral ``quarks" and
``antiquarks" in the fundamental representation of the $U(N_{f})$.
In the brane configuration where D8 and $\overline{\text{D8}}$ are
parallel, open-string excitations on each stack of branes transform
independently under the chiral transformation and thus we have a
chiral symmetric background.  The dual gauge matter will be in the
chiral symmetric phase.  On the other hand, in the connecting brane
configuration, chiral symmetry is broken at the tip and the
corresponding gauge matter will be in the chirally broken
phase~\cite{asy}.

To add the baryonic density to the boundary gauge matter, the
non-normalizable mode of the $a_{0}^{V}$ component of the
$U(1)\subset U(N_{f})$ field is turned on.  The baryon chemical
potential $\mu$ of the corresponding gauge matter is identified with
the non-normalizable mode of the DBI gauge field at the boundary
by~\cite{ksz}
\begin{eqnarray}
\mu & = & a^{V}_{0}(u\to\infty).
\end{eqnarray}

To turn on the magnetic field, another component $a_{3}^{V}$ is used
as the vector potential generating the magnetic field. The direction
of the magnetic field is chosen so that the vector potential is
\begin{eqnarray}
a^{V}_{3}& = & B x_{2}.
\end{eqnarray}
The Chern-Simon action in the background couples these two
components to the third component $a_{1}^{A}$ of the $U(1)$,
generating the response to the external magnetic field.  The
response appears as the gradient of the chiral condensate along the
direction of $B$ at the boundary which is defined to be
$a_{1}^{A}(u\to \infty)\equiv\mathcal{5}\varphi$.  Here and
henceforth, we will call $\mathcal{5}\varphi$ a pion gradient.

The DBI and the Chern-Simon actions are then given by
\begin{eqnarray}
S_{D8}& = & \mathcal{N}
\int^{\infty}_{u_{c}}du~u^{5/2}\sqrt{1+\frac{B^{2}}{u^{3}}}\sqrt{1+f(u)(a_{1}^{\prime
A})^{2}-(a_{0}^{\prime V})^{2}+f(u)u^{3}x_{4}^{\prime 2}},  \label{action1}  \\
S_{CS}& = & -\frac{3}{2}\mathcal{N}
\int^{\infty}_{u_{c}}du~(\partial_{2}a^{V}_{3}a^{V}_{0}a^{A
\prime}_{1}-\partial_{2}a^{V}_{3}a^{V \prime}_{0}a^{A}_{1}),
\label{action2}
\end{eqnarray}
where $\mathcal{N}=N_{c}R^{2}_{D4}/(6\pi^2(2\pi
\alpha^{\prime})^{3})$ defines the brane tension.  The factor $3/2$
in the Chern-Simon action comes from addition of surface term in
order to maintain the gauge invariance of the total action in the
situation when the gauge transformation does not vanish at the
boundary~(see Ref.~\cite{BLLm} for details).

Consequently, the equations of motion with respect to each gauge
field $a_{0}^{V},a_{1}^{A}$ are
\begin{eqnarray}
\frac{\sqrt{u^{5}+B^{2}u^{2}}~f(u)a_{1}^{\prime
A}}{\sqrt{1+f(u)(a_{1}^{\prime A})^{2}-(a_{0}^{\prime
V})^{2}+f(u)u^{3}x_{4}^{\prime 2}}}& = & j_{A}-\frac{3}{2}B\mu+3B
a_{0}^{V}, \label{eq:a0} \\
\frac{\sqrt{u^{5}+B^{2}u^{2}}~a_{0}^{\prime
V}}{\sqrt{1+f(u)(a_{1}^{\prime A})^{2}-(a_{0}^{\prime
V})^{2}+f(u)u^{3}x_{4}^{\prime 2}}}& = &
d-\frac{3}{2}Ba_{1}^{A}(\infty)+3B a_{1}^{A}. \label{eq:a1}
\end{eqnarray}
The corresponding density and current density, $d,j_{A}$, at the
boundary($u\to\infty$) are defined as
\begin{eqnarray}
j^{\mu}(x, u\to\infty)& \equiv & \frac{\delta S_{eom}}{\delta
A_{\mu}}\bigg{\vert}_{u\to\infty} \\
                      & \equiv & (d,\vec{j_{A}}).
\end{eqnarray}
They are related to the components of the $U(1)$ gauge field by
\begin{eqnarray}
d & = & \frac{\sqrt{u^{5}+B^{2}u^{2}}~a_{0}^{\prime
V}}{\sqrt{1+f(u)(a_{1}^{\prime A})^{2}-(a_{0}^{\prime
V})^{2}+f(u)u^{3}x_{4}^{\prime 2}}}\bigg{\vert}_{\infty}-\frac{3}{2}B a_{1}^{A}(\infty), \\
j_{A}& = & \frac{\sqrt{u^{5}+B^{2}u^{2}}~f(u)a_{1}^{\prime
A}}{\sqrt{1+f(u)(a_{1}^{\prime A})^{2}-(a_{0}^{\prime
V})^{2}+f(u)u^{3}x_{4}^{\prime
2}}}\bigg{\vert}_{\infty}-\frac{3}{2}B\mu.
\end{eqnarray}
For the phase of pure pion gradient where chiral symmetry is broken,
the axial current $j_{A}$ is set to zero and the density
$d=\frac{3}{2}B\mathcal{5}\varphi$ is the definition adapted from
the Wess-Zumino-Witten action of the boundary gauge
theory~\cite{sst}.

The constant of motion with respect to $x_{4}(u)$ for the pure pion
gradient phase yields
\begin{eqnarray}
(x^{\prime}_{4}(u))^{2}& = & \frac{1}{u^{3}f(u)}\Big[
\frac{u^{3}[f(u)(C(u)+D(u)^{2})-9B^{2}\Big(a_{0}^{V}-\frac{\mu}{2}\Big)^{2}]}{F^{2}}-1
\Big]^{-1},  \label{eq:x4prime}
\end{eqnarray}
where
\begin{eqnarray}
F & = & \frac{u^{3}_{c}
\sqrt{f(u_{c})}\sqrt{f(u_{c})(C(u_{c})+D(u_{c})^{2})-9B^{2}\Big(a_{0}^{V}(u_{c})-\frac{\mu}{2}\Big)^{2}}~x_{4}^{\prime}(u_{c})}{\sqrt{1+f(u_{c})u^{3}_{c}~x_{4}^{\prime
2}(u_{c})}}  \label{eq:F}  \\
& = & u_{c}^{3/2}\sqrt{f(u_{c})C(u_{c})-9B^{2}\left(
a_{0}^{V}(u_{c})-\frac{\mu}{2} \right)^{2}},
\end{eqnarray}
where $C(u)\equiv u^{5}+B^{2}u^{2}, D(u)\equiv
d-3B\mathcal{5}\varphi/2+3Ba_{1}^{A}(u)$.  $u_{c}$ is the position
where the D8 and $\overline{\text{D8}}$ branes connect.  Since there
is no instanton in this case, the branes connect smoothly at
$u_{c}$.
 We also have $D(u_{c})=0$ from $a_{1}^{A}(u_{c})=0$, and
 $x_{4}'(u_{c})=\infty$.

Since the DBI action, Eqn.~(\ref{action1}), is divergent from the
limit $u\to \infty$, we would need the action of the magnetized
vacuum for the regularization.  For the magnetized vacuum, we can
let the non-normalizable modes, $a^{V}_{0}, a^{A}_{1}=0$ and $d,
j_{A}=0$.  The vacuum action then takes the following form
\begin{eqnarray}
S[\text{magnetized vacuum}] & = &
\int^{\infty}_{u_{0}}~\sqrt{C(u)(1+f(u)u^{3}x^{\prime
2}_{4})}\bigg{\vert}_{vac}~du, \nonumber
\end{eqnarray}
where
\begin{eqnarray}
 x^{\prime}_{4}(u)\vert_{vac} & = &
\frac{1}{\sqrt{f(u)u^{3}\Big(\frac{f(u)u^{3}C(u)}{f(u_{0})u^{3}_{0}C(u_{0})}-1
\Big)}}.
\end{eqnarray}
Again, the position of the tip of the brane configuration is denoted
by $u_{0}$.  The temperature and field dependence of the position
$u_{0}$ are given in Fig.~1 of Ref.~\cite{pbm1}. It saturates
approximately at 1.23 for all temperatures at high magnetic field.
The action of the vacuum will be used to regulate the infinity of
the DBI action from the limit $u\to \infty$ when we calculate the
free energy of the dual gauge matter in the subsequent section.

\subsection{Zero Temperature case $f(u)=1$}

We can numerically solve the equations of motion,
Eqn.~(\ref{eq:a0}),(\ref{eq:a1}) by using the shooting algorithm.
However, it is illustrative to consider first the limiting case of
zero temperature where $f(u)=1$ and the equations of motion are
simplified so much that they yield exact analytic solutions. Later
on we will actually find from the numerical solutions that most
physical properties of the pion gradient phase are insensitive to
the change of temperature.

In the zero temperature case, starting from the equations of motion,
Eqn.~(\ref{eq:a0}),(\ref{eq:a1}), rewritten as
\begin{eqnarray}
a_{0}^{\prime V}(a_{0}^{V}-\frac{\mu}{2}) & = &
f(u)a_{1}^{A}a_{1}^{\prime A} \label{eom1}
\\
f(u)a_{0}^{\prime V}a_{1}^{\prime A} & = &
\frac{9B^{2}a_{1}^{A}(a_{0}^{V}-\frac{\mu}{2})}{u^{5}+B^{2}u^{2}}(1+fu^{3}x_{4}^{\prime
2}+f a_{1}^{\prime A 2}-a_{0}^{\prime V 2}). \label{eom2}
\end{eqnarray}
From Eqn.~(\ref{eom1}), for $f=1$ we can solve to obtain
\begin{eqnarray}
(a_{0}^{V}-\frac{\mu}{2})^{2}-(\frac{\mu}{2})^{2} & = & a_{1}^{A
2}-(\mathcal{5}\varphi)^{2}.  \label{12rel}
\end{eqnarray}
Using Eqn.~(\ref{eom2}), direct integration leads to
\begin{eqnarray}
a_{0}^{V} & = &
\frac{\mu}{2}+\sqrt{(\frac{\mu}{2})^{2}-(\mathcal{5}\varphi)^{2}}~\cosh
I(u) \label{sola0} \\
a_{1}^{A} & = & \sqrt{(\frac{\mu}{2})^{2}-(\mathcal{5}\varphi)^{2}}
~\sinh I(u) \label{sola1},
\end{eqnarray}
where
\begin{eqnarray}
I(u) & \equiv & \int_{u_{c}}^{u}~du
\sqrt{g(u,u_{c},B)\left[\frac{C(u)}{9B^{2}}-\left(
(\frac{\mu}{2})^{2}-(\mathcal{5}\varphi)^{2}\right)\right]^{-1}}, \\
g(u,u_{c},B) & \equiv & 1+ u^{3}x_{4}^{\prime 2}  \\
& = & 1+\left[
\frac{u^{3}(C(u)-9B^{2}((\frac{\mu}{2})^{2}-(\mathcal{5}\varphi)^{2}))}{u_{c}^{3}(C(u_{c})-9B^{2}((\frac{\mu}{2})^{2}-(\mathcal{5}\varphi)^{2}))}-1
\right]^{-1},
\end{eqnarray}
by using the boundary conditions $a_{1}^{A}(u_{c})=0$ and
Eqn.~(\ref{12rel}) at $u_{c}$.  Additionally, there are two
constraints which need to be satisfied,
\begin{eqnarray}
\cosh I_{\infty} & = &
\frac{\mu}{2}\frac{1}{\sqrt{(\frac{\mu}{2})^{2}-(\mathcal{5}\varphi)^{2}}},
\label{con1}   \\
1 & = & 2 \int_{u_{c}}^{\infty}  \label{con2}
du~x_{4}^{\prime}(u),
\end{eqnarray}
where $I_{\infty}\equiv I(u\to \infty)$.  In the zero temperature
case $x_{4}^{\prime}$ is given by
\begin{eqnarray}
x_{4}^{\prime}(u) & = & \left[ u^{3}\left(
\frac{u^{3}(C(u)-9B^{2}((\frac{\mu}{2})^{2}-(\mathcal{5}\varphi)^{2}))}{u_{c}^{3}(C(u_{c})-9B^{2}((\frac{\mu}{2})^{2}-(\mathcal{5}\varphi)^{2}))}-1
\right) \right]^{-1/2}.
\end{eqnarray}
The pion gradient is thus
\begin{eqnarray}
\mathcal{5}\varphi & = & \frac{\mu}{2}\tanh I_{\infty}.
\end{eqnarray}
In order to obtain the solutions, we numerically solve for $u_{c}$
from the constraints Eqn.~(\ref{con1}) and Eqn.~(\ref{con2})
simultaneously by fixing two parameters among
$(B,d,\mathcal{5}\varphi,\mu)$. The solutions always have
$\frac{\mu}{2}>\mathcal{5}\varphi$ as a reality condition.

As is found in Ref.~\cite{prs} for the pure pion gradient and
Ref.~\cite{bll,pbm} for the model with instantons, there are 2
possible brane configurations satisfying the scale fixing condition
$L_{0}=1$, one with small and one with large $u_{c}$.  The brane
configuration with small $u_{c}$ has longer stretch in the
$u$-direction and therefore has higher energy than the configuration
with large $u_{c}$.  The excess energy makes this configuration less
preferred thermodynamically.  For the pure pion gradient phase,
there is also the small-$u_{c}$ configuration~(for sufficiently
large $\mu$ and small $B$) which we found to be less preferred
thermodynamically even than the vacuum. Therefore this configuration
will not be considered in this article.

\begin{figure}[htp]
\centering
\includegraphics[width=0.45\textwidth]{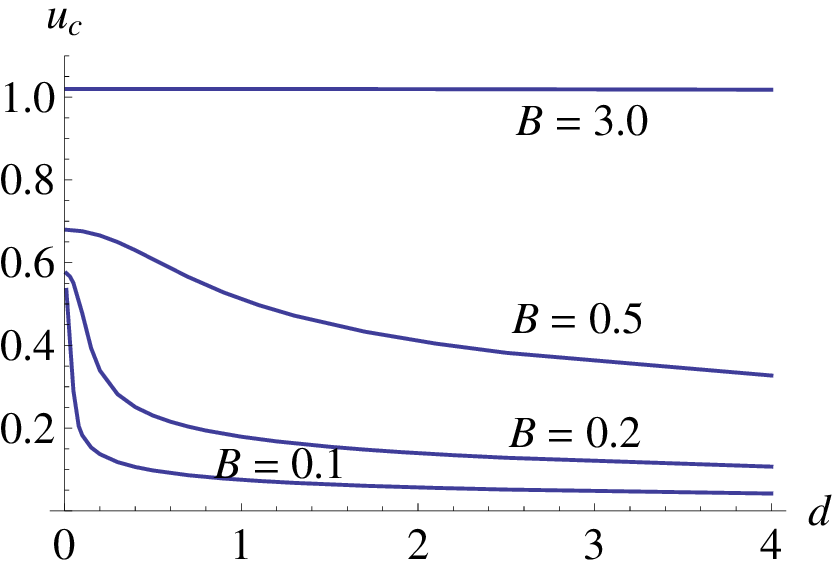} \hfill
\includegraphics[width=0.45\textwidth]{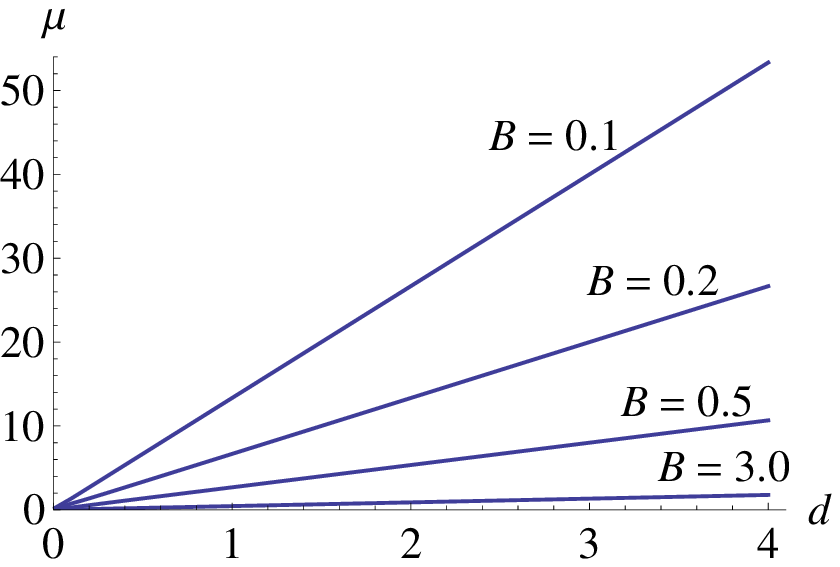}\\
\caption[]{The position $u_{c}$ (a) and the chemical potential (b)
as a function of the density at fixed magnetic field $B$ at zero
temperature. } \label{fig1.0}
\end{figure}

The solutions can be explored by slicing through the plane in the
parameter space at fixed magnetic field~($B$), fixed density~($d$),
and fixed chemical potential~($\mu$) respectively.  At fixed $B$,
the solutions are shown in Fig.~\ref{fig1.0} for the position
$u_{c}$ and the chemical potential as a function of the density. For
small $B$, the position $u_{c}$ has certain variation with respect
to the density.  As $B$ increases, $u_{c}$ saturates to an almost
constant curve with a slight density dependence.  The chemical
potential at fixed $B(\geq 0.1)$ is found to be an exact linear
function of the density.  The slope of the linear function is
inversely proportional to $B$.  The relation can be summarized into
the following simple form
\begin{eqnarray}
\mu & = & \frac{4}{3}\frac{d}{B} ~~~~~\text{  for }B \geq 0.1.
\label{eqmud}
\end{eqnarray}
This implies from $d = 3B \mathcal{5}\varphi/2$ that $\mu =
2\mathcal{5}\varphi$ for $B \geq 0.1$.  The linear relation between
$\mu$ and $d$ can be interpreted as the absence of self-interaction
among the pion gradient excitations.  Each pion gradient excitation
seems to behave as free entity for $B \geq 0.1$.

\begin{figure}[htp]
\centering
\includegraphics[width=0.65\textwidth]{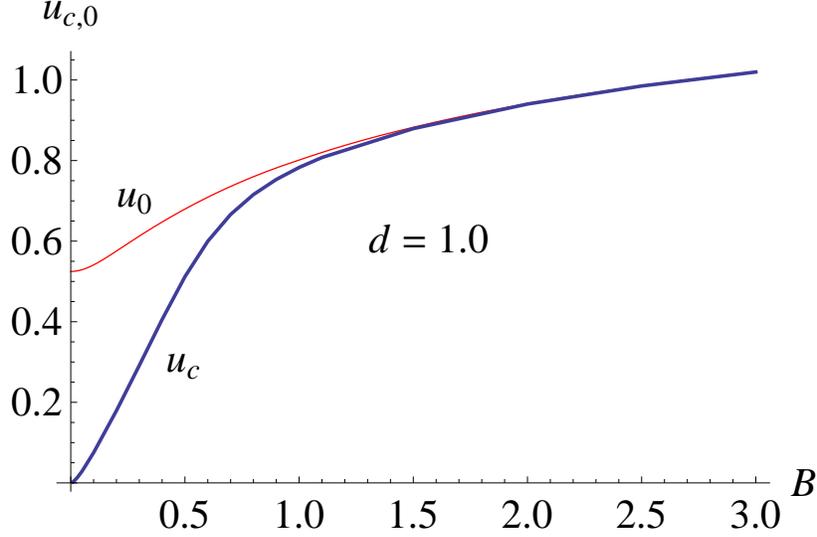}
\caption[]{Position $u_{c}$ as a function of $B$ at a fixed density
$d=1.0$ at zero temperature, the position $u_{0}$ of the magnetized
vacuum is shown for comparison. } \label{fig1.1}
\end{figure}

\begin{figure}[htp]
\centering
\includegraphics[width=0.65\textwidth]{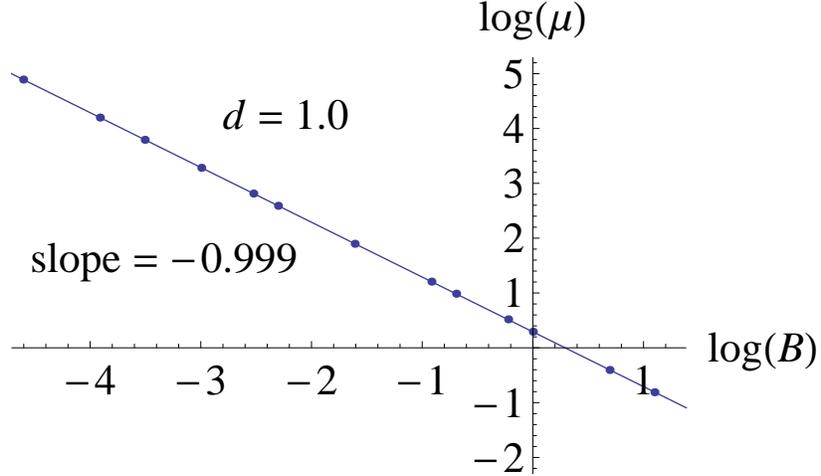}
\caption[]{Chemical potential as a function of $B$ at $d=1.0, T=0$
in the logarithmic scale. } \label{fig1.2}
\end{figure}

For fixed $d$, the position $u_{c}$ and the chemical potential are
shown as functions of $B$ in Fig.~\ref{fig1.1},\ref{fig1.2} for
$d=1.0$. Solutions exist for the entire range of $B$, down to
$u_{c}(B=0)=0$.  Consistent with what previously found, $\mu$ is
found to be inversely proportional to $B$ as is shown in
Fig.~\ref{fig1.2}.

\begin{figure}[htp]
\centering
\includegraphics[width=0.45\textwidth]{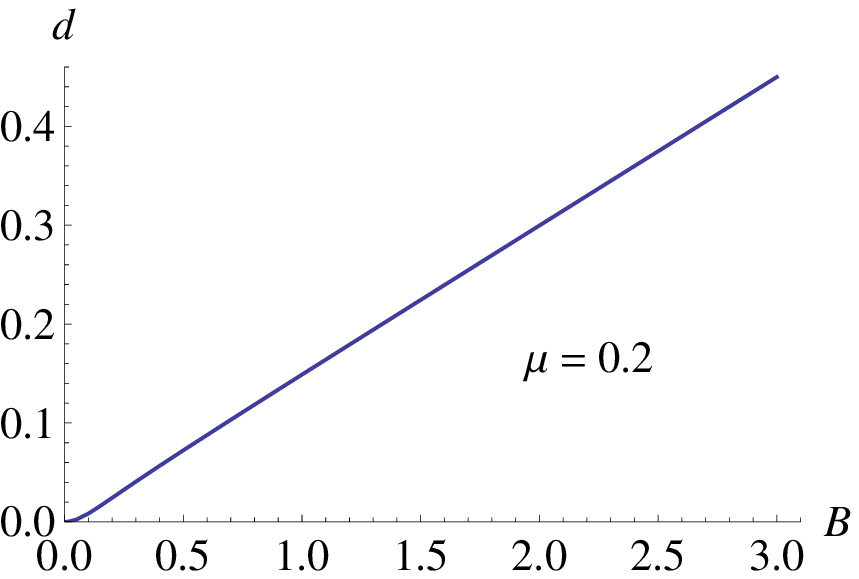} \hfill
\includegraphics[width=0.45\textwidth]{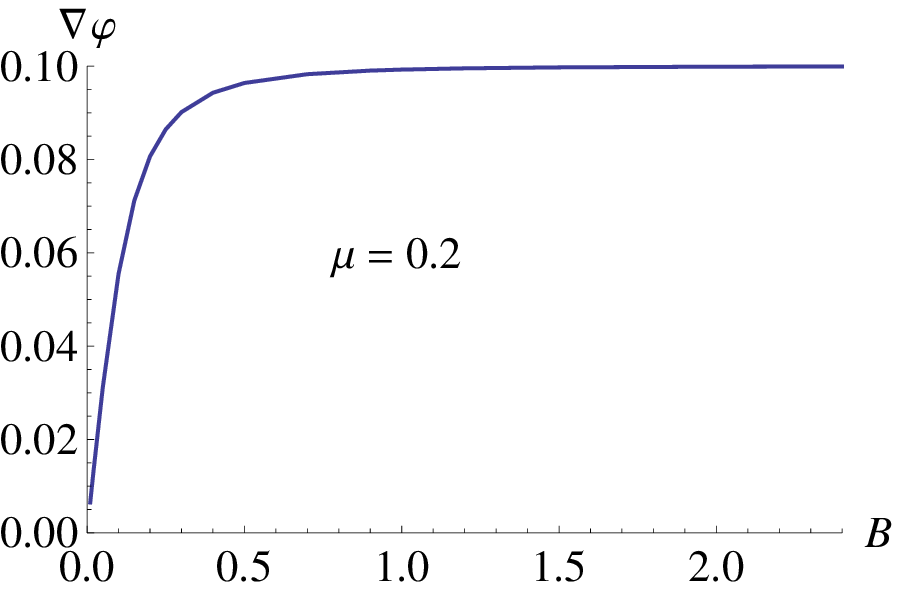} \\
\includegraphics[width=0.45\textwidth]{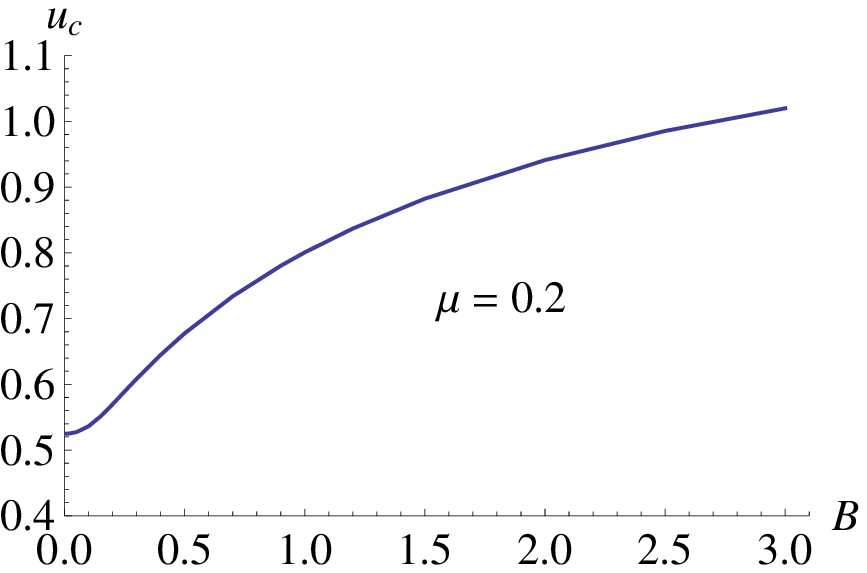} \\
\caption[]{The density, the pion gradient, and the position $u_{c}$
as a function of $B$ at fixed $\mu=0.2, T=0$. } \label{fig1.3}
\end{figure}

\begin{figure}[htp]
\centering
\includegraphics[width=0.65\textwidth]{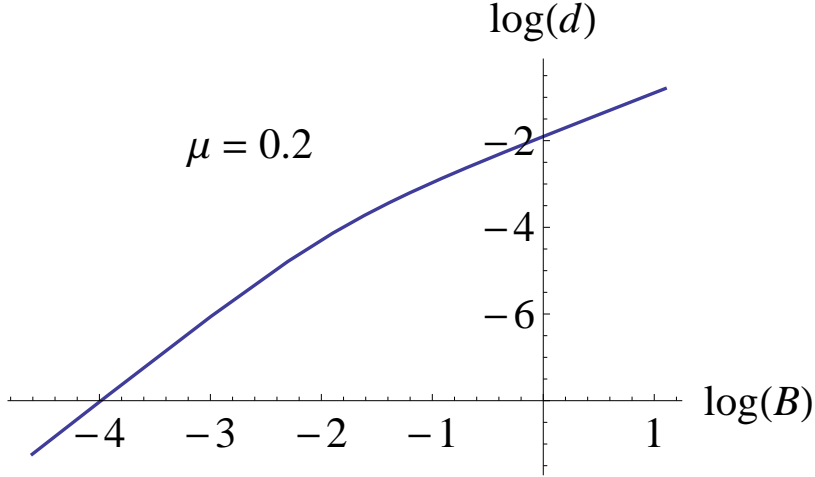}
\caption[]{The density as a function of $B$ at $\mu=0.2, T=0$ in the
logarithmic scale. } \label{fig1.31}
\end{figure}

For fixed $\mu$, Fig.~\ref{fig1.3} shows interesting transition
between 2 regions of the parameter space.  In Fig.~\ref{fig1.31},
the relation between $d$ and $B$ is shown to be approximately
quadratic for $B\leq 0.2$ and linear for $B\gtrsim 0.2$.  From $d=
3B\mathcal{5}\varphi/2$, this implies that $\mathcal{5}\varphi$ is a
linear function of $B$ for $B\leq 0.2$ and a constant function for
$B\gtrsim 0.2$. Fig.~\ref{fig1.3}(b) confirms the behaviour.  Since
the saturation at large $B$ occurs around
$\mathcal{5}\varphi=\mu/2$, the slope of the linear region, $B\leq
0.2$, is therefore proportional to $\mu$. Consequently, for small
$B$, $\mathcal{5}\varphi \sim \mu B$. The behaviour at small $B$ is
similar to the behaviour found in Ref.~\cite{sst,BLLm} for the
confined phase.  The result in the deconfined phase of the SS model
at zero temperature was first obtained in Ref.~\cite{prs}.

It should be noted that the nonlinear effects of the DBI action
become apparent for $B\gtrsim 0.2$ where $\mathcal{5}\varphi \simeq
\mu/2$.  As $B$ increases, the pion gradient does not change but its
baryonic density increases linearly with the field.  This
$\mathcal{5}\varphi$-saturation is a new effect observed only in the
theory with DBI gauge interaction.

\section{Thermodynamical Properties of the Pure Pion Gradient Phase}

In the pure pion gradient phase, since $x_{4}^{\prime}\to \infty$ at
$u_{c}$, the integrand of the action diverges at $u_{c}$ in addition
to the limit $u\to \infty$.  This also occurs with the magnetized
vacuum where $x_{4}^{\prime}$ is divergent at $u_{0}$.  However, the
limit which makes the integral and consequently the action divergent
comes only from $u\to \infty$~(the divergences at $u_{0,c}$ are
weaker than a simple pole and thus finite over integration). We can
therefore regulate the action by subtracting the total action with
the action of the magnetized vacuum in the usual manner.

The regulated free energy is thus given by
\begin{eqnarray}
\mathcal{F}_{E}(d,B)=\Omega(\mu,B) +\mu d,
\end{eqnarray}
where $\Omega(\mu,B) =
S[a_{0}(u),a_{1}(u)](e.o.m.)-S[\text{magnetized vacuum}]\equiv
\mathcal{F}(\mu,B)$.

\begin{figure}[htp]
\centering
\includegraphics[width=0.65\textwidth]{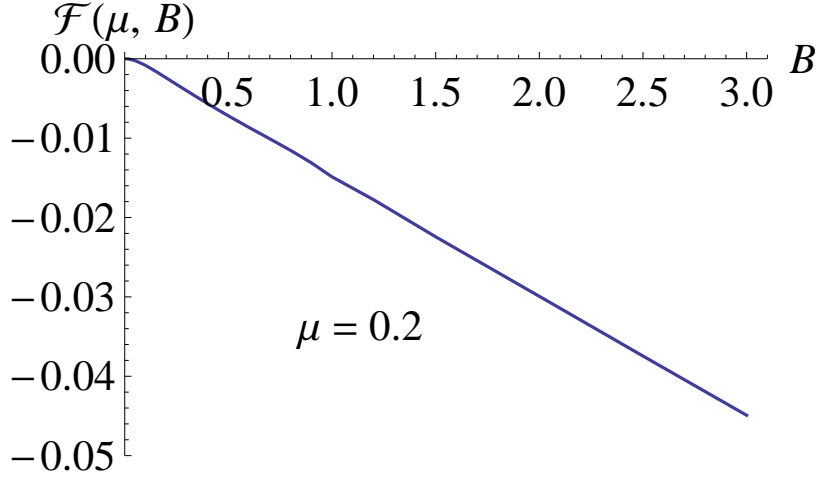}
\caption[]{The free energy as a function of $B$ at fixed $\mu=0.2,
T=0$. } \label{fig1.32}
\end{figure}

\begin{figure}[htp]
\centering
\includegraphics[width=0.65\textwidth]{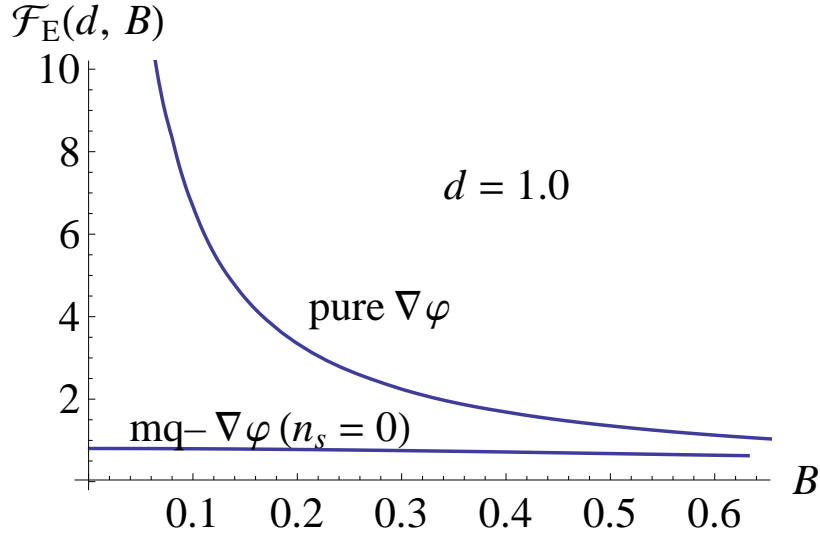}
\caption[]{The free energy as a function of the density of the pure
pion gradient phase compared to the multiquark-$\mathcal{5}\varphi$
phase at fixed $d=1.0, T=0$. } \label{fig1.5}
\end{figure}

We can calculate the total action satisfying the equation of motion
$S[a_{0}(u),a_{1}(u)](e.o.m.)= S_{D8}+S_{CS}$ to be
\begin{eqnarray}
S_{D8}& = &\mathcal{N}
\int^{\infty}_{u_{c}}du~C(u)\sqrt{\frac{f(u)(1+f(u)u^{3}{x^{\prime
2}_{4})}}{f(u)(C(u)+D(u)^{2})-9B^{2}\Big(a_{0}^{V}-\frac{\mu}{2}\Big)^{2}}}, \\
S_{CS}& = & -\mathcal{N} \frac{3}{2}B
\int^{\infty}_{u_{c}}du~\frac{\Big(
3Ba^{V}_{0}(a_{0}^{V}-\frac{\mu}{2})-f(u)
D(u)a^{A}_{1}\Big)\sqrt{\frac{1}{f(u)}+u^{3}x_{4}^{\prime
2}}}{\sqrt{f(u)(C(u)+D(u)^{2})-9B^{2}\Big(a_{0}^{V}-\frac{\mu}{2}\Big)^{2}}}.
\label{CSL}
\end{eqnarray}
For zero temperature the total action reduces to
\begin{eqnarray}
S_{e.o.m.} & = & \mathcal{N} \int_{u_{c}}^{\infty}~du
\sqrt{\frac{g(u,u_{c},B)}{C(u)-9B^{2}((\frac{\mu}{2})^{2}-(\mathcal{5}\varphi)^{2})}}\left(
C(u)-\frac{9B^{2}}{2}(\frac{\mu}{2}a_{0}^{V}(u)-(\mathcal{5}\varphi)^{2})
\right). \nonumber
\end{eqnarray}
We can compute this action by substituting Eqn.~(\ref{sola0}) into
the expression.  The free energy at fixed chemical potential
$\mathcal{F}(\mu,B)$ of the pure pion gradient phase at zero
temperature is shown in Fig.~\ref{fig1.32}.  Once $d,\mu > 0$, the
free energy becomes smaller than the free energy of the magnetized
vacuum~(being negative) and thus thermodynamically preferred.  The
magnetization at fixed chemical potential
$\mathcal{M}(\mu,B)=-\frac{\partial \mathcal{F}(\mu,B)}{\partial B}$
therefore increases from zero and becomes constant
$M(\mu=0.2,B)\simeq 0.0152$ at large field~($B> 0.2$) as we can see
from the slope of Fig.~\ref{fig1.32}.  On the other hand, the
magnetization at fixed $d=1.0$, $M(d,B)= -\frac{\partial
\mathcal{F}_{E}(d,B)}{\partial B}$, of the pure pion gradient phase
is a rapidly decreasing function of $B$ as is shown in
Fig.~\ref{fig1.5}.

The pressure of the pure pion gradient as a function of the density
can be calculated using Eqn.~(\ref{eqmud}) and $d=\frac{\partial
P}{\partial \mu}$~(see Ref.~\cite{bhp}),
\begin{eqnarray}
P(d,B) & = & \mu(d,B)d-\int_{0}^{d}\mu(d',B)~d(d'),  \\
& = & \frac{1}{2}k(B)d^{2},  \label{eqpd}
\end{eqnarray}
where $k(B)=4/3B$ for $B\geq 0.1$.  The quadratic dependence of the
pressure on the density without higher order term reveals that the
pion gradient excitations behave like free particles without either
repulsive or attractive interaction among themselves.

For the parameter space in the region $d\ll 1, B\ll 1$ such as the
regions shown in Fig.~\ref{fig1.3}, \ref{fig1.31}, since
$\mathcal{5}\varphi \sim \mu B, d=3B\mathcal{5}\varphi/2$, we have
$d=\alpha \mu B^{2}$ for some constant $\alpha$.  In this case, the
linear relations between $\mu$ and $d$ is still valid and the
equation of state is again given by Eqn.~(\ref{eqpd}) with
$k(B)=1/\alpha B^{2}$~(for $\mu=0.2, \alpha \simeq 4.634$).  This
behaviour is similar to what found in Ref.~\cite{sst} using the
Wess-Zumino-Witten term in the boundary theory and in
Ref.~\cite{BLLm,prs} for the confined and deconfined SS model at
zero temperature.

The energy density can be calculated straightforwardly
\begin{eqnarray}
\rho & = & \int_{0}^{d}~\mu(\eta,B)~d\eta, \\
& = & \frac{1}{2}k(B)d^{2},
\end{eqnarray}
where $k(B)=1/\alpha B^{2}, 4/3B$ for small and large $B$
respectively.  The results are remarkably similar to the results
from the bottom-up AdS/QCD model considered in Ref.~\cite{st}. The
equation of state then becomes simply $P=\rho$ representing free gas
of the solitonic excitations of the the pion gradient. The adiabatic
index,~$\Gamma$, and the sound speed,~$c_{s}$, are then calculated
to be
\begin{eqnarray}
\Gamma & \equiv & \frac{\rho}{P}\frac{\partial P}{\partial
\rho}=\frac{\rho}{P}c_{s}^{2}=1, \\
c_{s} & = & 1,
\end{eqnarray}
the typical behaviour of the free gas.

\begin{figure}[htp]
\centering
\includegraphics[width=0.45\textwidth]{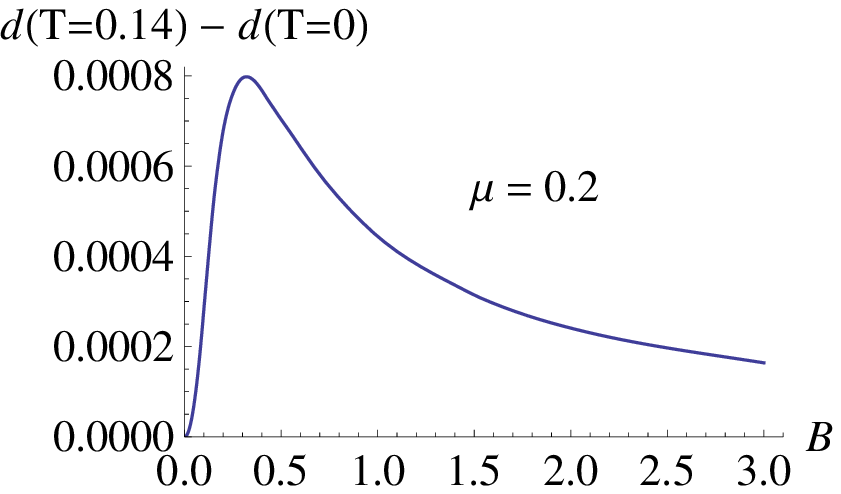} \hfill
\includegraphics[width=0.5\textwidth]{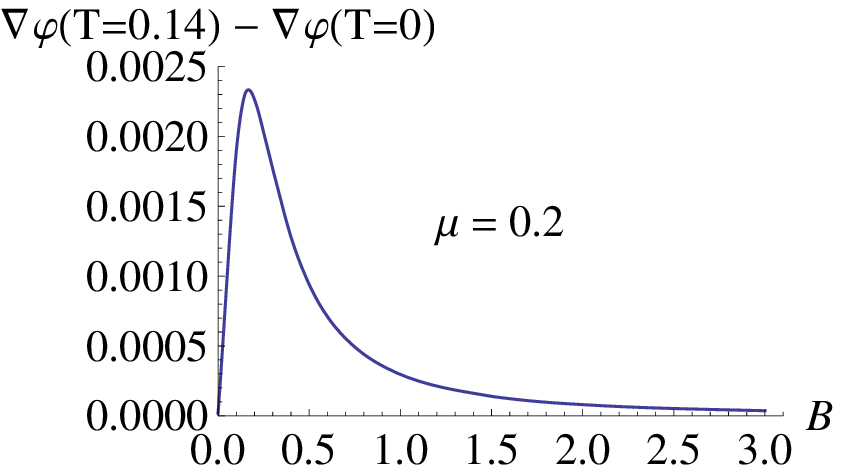} \\
\includegraphics[width=0.45\textwidth]{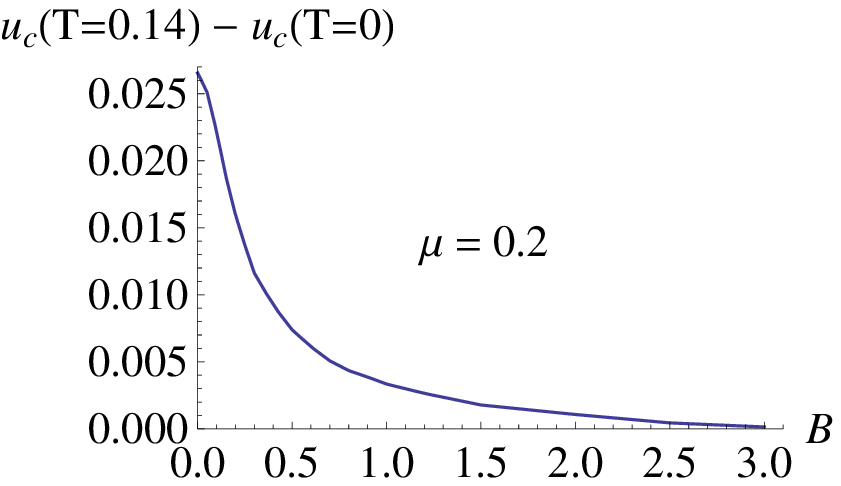} \\
\caption[]{The difference between the density, the pion gradient,
and the position $u_{c}$ of the $\mathcal{5}\varphi$ phase at
$T=0.14$ and $T=0$ as a function of $B$ for $\mu = 0.2$. }
\label{fig1.4}
\end{figure}

For nonzero temperature, the full equations of motion,
Eqn.~(\ref{eq:a0}),(\ref{eq:a1}) can be solved numerically by double
shooting algorithm aiming for two conditions to be satisfied at
once: $a_{1}^{A}(u_{c})=0, L_{0}=1$ while fixing $B,\mu$ and
$d$~(and consequently $\mathcal{5}\varphi$). It is found that the
temperature dependence of every physical quantity of the pure pion
gradient is very weak. Figure~\ref{fig1.4} shows the {\it
difference} of the density, the pion gradient, and the position
$u_{c}$ at fixed $\mu = 0.2$ between $T=0.14$ and $T=0$. Observe
that the difference in the temperature dependence of the density and
the pion gradient are the most distinctive in the transition region
when the magnetic field changes from small to large values.

\subsection{Comparison to the Multiquark-$\mathcal{5}\varphi$ phase}

We would like to consider whether the pure pion gradient phase is
thermodynamically preferred than the other possible nuclear phases
exist in the same parameter space.  In the deconfined SS model in
the presence of the magnetic field, there are generically 3 possible
phases in addition to the vacuum; the chiral symmetric QGP phase,
the chirally broken phase of
multiquark-$\mathcal{5}\varphi$~(MQ-$\mathcal{5}\varphi$), and the
pure pion gradient~($\mathcal{5}\varphi$) phase.  The multiquark
nuclear phase has been studied in Ref.~\cite{bch,pbm,pbm1} and found
to be the most preferred phase for the dense deconfined nuclear
matter under moderate external magnetic fields.  The multiquark
phase actually has certain mixture of the pion gradient as the
source for the baryon density.  This is inevitable since the
response of the nuclear matter to the external magnetic field is in
the form of the spatial variation of the chiral condensate in the
direction of the applied field which we call the pion gradient.

However, the ratio of the pion gradient population with respect to
the multiquark decreases as $d$ grows~\cite{pbm1}. It is thus
suggestive that the multiquark phase is likely to be more
thermodynamically preferred than the pion gradient phase.  In this
subsection we directly compare the two phases at zero temperature
using the free energy at fixed density $d=1.0$.  The
MQ-$\mathcal{5}\varphi$ phase imposes the boundary conditions~(see
Ref.~\cite{pbm,pbm1} for details);
\begin{equation*}
j_{A}=0,~a_{1}^{A}(u_{c})=0,
~a_{0}^{V}(u_{c})=\frac{1}{3}u_{c}\sqrt{f(u_{c})}+n_{s}(u_{c}-u_{T}),
\end{equation*}
where $n_{s}$ is the number of colour strings~(hanging from the
baryon vertex down to the horizon) in fractions of $1/N_{c}$.

The result is shown in Fig.~\ref{fig1.5}. Clearly, the
multiquark-pion gradient~(MQ-$\mathcal{5}\varphi$) phase is more
preferred than the pure pion gradient phase.  Similar behaviours are
confirmed for small $d\simeq 0.1$ and large $d\gg 1$.  Especially at
large densities, since the baryon chemical potential of the
MQ-$\mathcal{5}\varphi$ phase increases slower than a linear
function~\cite{pbm} whilst it is linear for the $\mathcal{5}\varphi$
phase, the dominant term $\mu d$ in the free energy for the
MQ-$\mathcal{5}\varphi$ phase becomes much smaller and thus more
stable thermodynamically.

\begin{figure}[htp]
\centering
\includegraphics[width=0.65\textwidth]{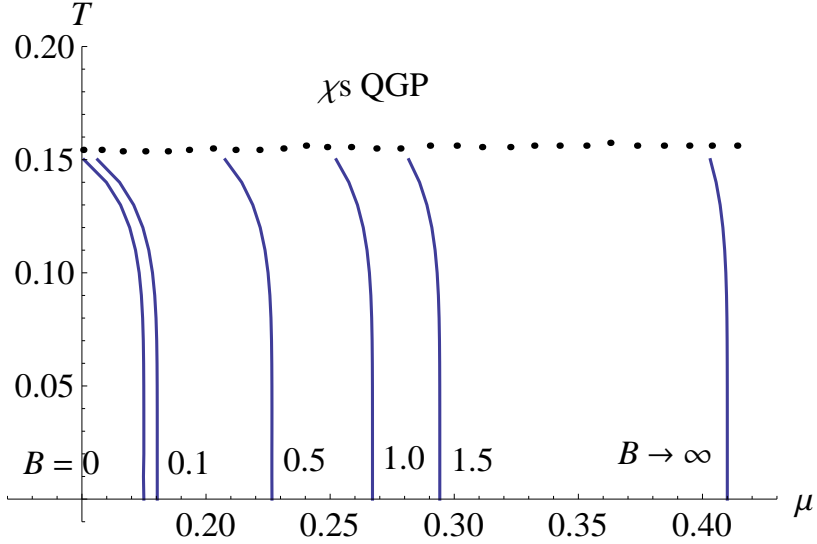}
\caption[]{The onset chemical potential of the
multiquark-$\mathcal{5}\varphi$ phase as a function of $T,B$~(for
$B\to \infty, u_{0}=1.23$ is used). These lines can be served as the
transition lines between the $\mathcal{5}\varphi$ phase on the left
and multiquark-$\mathcal{5}\varphi$ phase~($n_{s}=0$) on the right.
The dotted line represents schematic transition to the chiral
symmetric QGP phase. } \label{fig1.6}
\end{figure}

However, there is a region of parameter space where the
$\mathcal{5}\varphi$ phase is dominant. When the baryon chemical
potential $\mu < \mu_{\text{onset}}\equiv
\mu(d=0)=\frac{1}{3}u_{0}\sqrt{f(u_{0})}+n_{s}(u_{0}-u_{T})$ of the
multiquark, the multiquarks cease to exist and the pion gradient
which can be constructed at arbitrarily small $\mu$~(since $\mu \sim
d$) will be dominating.  The corresponding transition line in the
$(\mu,T)$ diagram for $B=0$ is shown in Fig.~8 of Ref.~\cite{bch}.
For $B>0$, dependence of $u_{0}$ on $B$ affects the transition line
accordingly as shown in Fig.~\ref{fig1.6}.  The dotted line
represents schematic transition to the chiral symmetric quark-gluon
plasma~($\chi S$-QGP) phase.  The chiral symmetry restoration
between the magnetized vacuum and the $\chi S$-QGP has been studied
in Ref.~\cite{jk}.  The transition between the pure pion gradient
phase and the $\chi S$-QGP has been explored in Ref.~\cite{prs} with
$f=1$ approximation for the pure pion gradient. Since we found that
the $\mathcal{5}\varphi$ phase is insensitive to the change of
temperature, the results in Ref.~\cite{prs} should be justified to
be a good approximated phase diagram.  The chiral symmetry
restoration between the MQ-$\mathcal{5}\varphi$ phase and the $\chi
S$-QGP phase has been investigated in Ref.~\cite{pbm1}.

\section{Conclusions and Discussions}

The behaviour of the chirally broken pure pion gradient phase in the
deconfined SS model is studied at zero temperature and subsequently
finite temperature.  The magnetic response of the chirally broken
phase is linear $\mathcal{5}\varphi \sim \mu B$ for small field and
saturates to constant value $\mathcal{5}\varphi \sim \mu/2$ for
large field.

Relationship between $\mu$ and $d$ is also linear $\mu = k(B)d$
where $k(B)\sim 1/B^{2}, 1/B$ for small and large $B$ respectively.
This implies that the excitations of the pion gradient behave like a
free gas with no interaction among each other.  The equation of
state is thus simply $P=\rho$ with the sound speed equal to the
speed of light.  The free energies at fixed $\mu$ and $d$ are
obtained numerically.  Magnetization at fixed $\mu$ increases with
$B$ for small field and drops to constant value for large field.
Magnetization at fixed $d$ is a decreasing function with respect to
the magnetic field.

Using the free energy at fixed density, we show that the pure
$\mathcal{5}\varphi$ phase is less preferred thermodynamically than
the MQ-$\mathcal{5}\varphi$ phase at zero temperature.  The
configuration of the pure pion gradient phase is found to be
insensitive to the change of temperature, the difference of the free
energy at fixed $\mu$ for $T=0$ and $T=0.14$ is minimal, only about
$\lesssim 2\times 10^{-4}$.  On the other hand, the free energy of
the MQ-$\mathcal{5}\varphi$ phase is a decreasing function in the
temperature~\cite{pbm1}.  Therefore, we can conclude that the pure
pion gradient phase is generically less preferred than the
MQ-$\mathcal{5}\varphi$ for general situation.

However, there is an exception for small chemical potential, $\mu
\simeq 0.175-0.41$. When $\mu<\mu_{\text{onset}}$ of the
multiquarks, the multiquarks simply cannot exist while the pion
gradient can be induced at arbitrarily small $\mu$.  Therefore, in
this region of the parameter space, the pure pion gradient phase is
dominating over any other phases.  The transition lines are given by
$\mu=\mu_{\text{onset}}$ in the $(\mu,T)$ plane.  The interior of
certain classes of the dense astrophysical objects such as the
magnetars~\cite{dt} would have the corresponding regions where the
chemical potential~(and the density) and temperature fall into this
range.  In those regions, the dominating nuclear phase which governs
physics of the stars would be the pure pion gradient phase.

\section*{Acknowledgments}
\indent  I would like to thank CERN Theoretical Physics Division for
the warm hospitality during my visit where this work is completed.
P.B. is supported in part by the Thailand Research Fund~(TRF) and
Commission on Higher Education~(CHE) under grant RMU5380048 and
Thailand Center of Excellence in Physics~(ThEP).  P.B. and T.C. are
supported in part by the 90th Year Chulalongkorn Scholarship.

\newpage

\end{document}